\documentclass[journal=ancac3,manuscript=article]{achemso}

\usepackage{amsmath}

\author{Benjamin Lowe}
\affiliation{Institute of Physics, Academy of Sciences of the Czech Republic, Cukrovarnická 10, 162 00, Prague, Czech Republic}
\email{lowe@fzu.cz}

\author{Jan Berger}
\affiliation{Institute of Physics, Academy of Sciences of the Czech Republic, Cukrovarnická 10, 162 00, Prague, Czech Republic}
\altaffiliation
{Regional Centre of Advanced Technologies and Materials, Czech Advanced Technology and Research Institute (CATRIN), Palacký University Olomouc, \v{S}lechtitel\r{u} 241/27, 779 00 Olomouc, Czech Republic}

\author{Martin Ondr\'a\v{c}ek}
\affiliation{Institute of Physics, Academy of Sciences of the Czech Republic, Cukrovarnická 10, 162 00, Prague, Czech Republic}

\author{Oleksandr Stetsovych}
\affiliation{Institute of Physics, Academy of Sciences of the Czech Republic, Cukrovarnická 10, 162 00, Prague, Czech Republic}

\author{Shayan Louie}
\affiliation{Department of Chemistry, Columbia University, New York, NY, USA}

\author{Marco Lozano}
\affiliation{Institute of Physics, Academy of Sciences of the Czech Republic, Cukrovarnická 10, 162 00, Prague, Czech Republic}

\author{Diego Soler-Polo}
\affiliation{Institute of Physics, Academy of Sciences of the Czech Republic, Cukrovarnická 10, 162 00, Prague, Czech Republic}
\altaffiliation{IMDEA Nanoscience, Madrid 28049, Spain}

\author{Shiladitya Karmarkar}
\affiliation{Department of Chemistry, School of Sciences and Humanities, SR University, Warangal, Telangana - 506371, India}
\alsoaffiliation{Institute of Physics, Academy of Sciences of the Czech Republic, Cukrovarnická 10, 162 00, Prague, Czech Republic}

\author{Andr\'es Pinar Sol\'e}
\affiliation{Institute of Physics, Academy of Sciences of the Czech Republic, Cukrovarnická 10, 162 00, Prague, Czech Republic}

\author{Iva Plutnarov\'a}
\affiliation{Department of Inorganic Chemistry, University of Chemistry and Technology Prague, Technická 5, 166 28 Prague 6, Czech Republic}

\author{Pingo Mutombo}
\affiliation{Institute of Physics, Academy of Sciences of the Czech Republic, Cukrovarnická 10, 162 00, Prague, Czech Republic}

\author{Manish Kumar}
\affiliation{Institute of Physics, Academy of Sciences of the Czech Republic, Cukrovarnická 10, 162 00, Prague, Czech Republic}

\author{Michael P\'iro}
\affiliation{Institute of Physics, Academy of Sciences of the Czech Republic, Cukrovarnická 10, 162 00, Prague, Czech Republic}

\author{Ali Cem Demirok}
\affiliation{Institute of Physics, Academy of Sciences of the Czech Republic, Cukrovarnická 10, 162 00, Prague, Czech Republic}

\author{Michael L. Steigerwald}
\affiliation{Department of Chemistry, Columbia University, New York, NY, USA}

\author{Zden\v{e}k Sofer}
\affiliation{Department of Inorganic Chemistry, University of Chemistry and Technology Prague, Technická 5, 166 28 Prague 6, Czech Republic}

\author{Colin Nuckolls}
\affiliation{Department of Chemistry, Columbia University, New York, NY, USA}

\author{Pavel Jel\'inek}
\affiliation{Institute of Physics, Academy of Sciences of the Czech Republic, Cukrovarnická 10, 162 00, Prague, Czech Republic}
\altaffiliation
{Regional Centre of Advanced Technologies and Materials, Czech Advanced Technology and Research Institute (CATRIN), Palacký University Olomouc, \v{S}lechtitel\r{u} 241/27, 779 00 Olomouc, Czech Republic}
\email{jelinekp@fzu.cz}

\title{Rationalizing charge transfer: a single-molecule case study across three distinct surfaces}

\begin{document}



\begin{abstract}
Organic molecules are versatile building blocks for the formation of complex low-dimensional nanostructures on surfaces. Recently, interest has grown in translating on-surface synthetic protocols from noble metal surfaces to insulating surfaces, which offer technological relevance, and to magnetic or superconducting surfaces, which offer new routes to correlated phases. It is important, therefore, to understand how these alternative surfaces may alter the properties of adsorbates through processes like charge transfer and screening. Evidence for charge transfer is often obtained experimentally via scanning tunneling microscopy (STM) and spectroscopy (STS), and theoretically via density functional theory (DFT). In this work, we systematically compare the properties of isolated macrocyclic emeraldine (Me$_6$) molecules on three different surfaces: metallic Au(111), a semiconducting ferromagnetic monolayer of FeCl$_2$, and a cleaved superconducting surface of bulk 2H-NbSe$_2$. We show that STS and DFT alone are insufficient to quantify charge transfer for these cases. Instead, we complement these methods with Kelvin probe force microscopy (KPFM) measurements and model Hamiltonian calculations to demonstrate that on Au(111) and FeCl$_2$, Me$_6$ molecules donate a single electron to the substrate, whereas on NbSe$_2$ two electrons are transferred from molecule to substrate. Our work provides insight into the interplay of charge transfer, screening, and molecule-substrate coupling, and develops experimental and theoretical methods for comparing the properties of adsorbates across different surfaces for future studies.

\end{abstract}

Molecules adsorbed on atomically well-defined metal surfaces 
provide versatile building blocks for the bottom-up fabrication 
of complex low-dimensional quantum materials with tailored 
structural and electronic properties. By exploiting a delicate 
interplay between intermolecular interactions, metal--organic 
coordination, and on-surface covalent chemistry, molecular 
precursors can be assembled into well-defined one-dimensional 
chains, two-dimensional networks, and extended molecular 
architectures with atomic-scale precision 
\cite{Barth2005,Bartels2010}.  Recent developments in on-surface synthesis of organic and metal-organic materials have unlocked new classes of magnetic materials.\cite{song_-surface_2021,hou_construction_2025} The ability to stabilize previously unobtainable radical products opened the field of $\pi$-magnetism\cite{pavlicek_synthesis_2017,mishra_topological_2020,mishra_synthesis_2025,frezza_-surface_2024} with recent developments including the synthesis of spin-chains,\cite{su_fabrication_2025,sun_-surface_2025,zhao_spin_2025} while developments in 2D metal-organic frameworks have also resulted in promising magnetic properties including ferromagnetism,\cite{lobo-checa_ferromagnetism_2024,isufi_neziri_2d_2025} and rich magnetic phases mediated by coupling with radical ligands.\cite{frezza_-surface_2025} Despite some recent progress,\cite{kinikar_atomic-scale_2025,lowe_local_2024,dominguez-celorrio_systematic_2025,pawlak_toward_2025,yan_two-dimensional_2021,edens_spin_2025,vaxevani_parity-dependent_2026} the necessary catalytic contribution from metal surfaces in the formation of many of these products has limited this field predominantly to noble metal surfaces. The ability to translate these on-surface protocols to insulating surfaces would enhance their technological prospects, and the ability to combine these materials with magnetic or superconducting surfaces may lead to new correlated states.

\begin{figure}
    \centering
    \includegraphics[width=\linewidth]{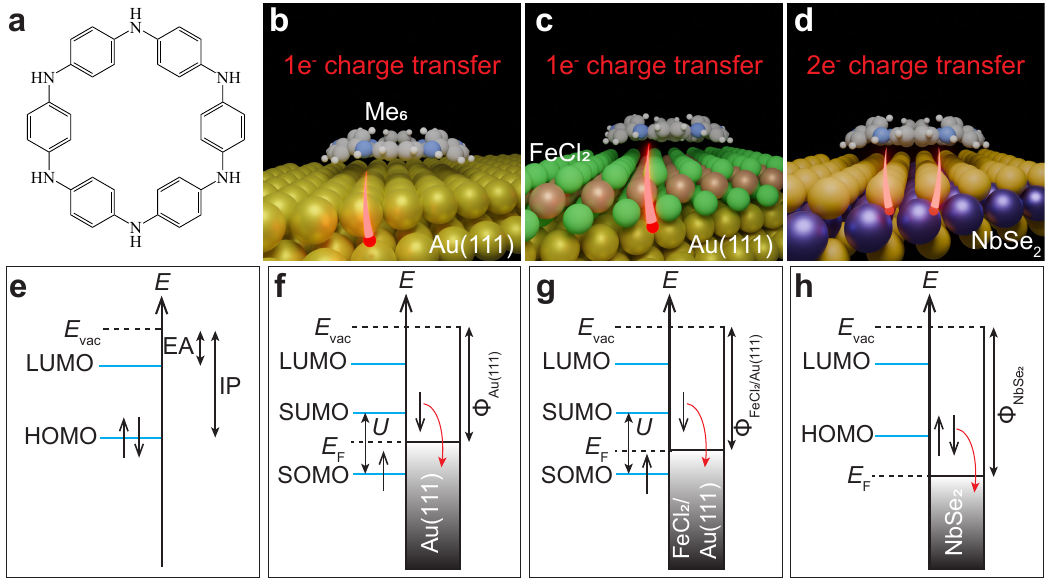}
    \caption{Charge transfer from Me$_6$ molecules to Au(111), FeCl$_2$/Au(111), and NbSe$_2$ substrates. \textbf{a} Chemical structure of Me$_6$. \textbf{b-d} Illustration of charge transfer from Me$_6$ to each of the three substrates Au(111), ML FeCl$_2$/Au(111), and NbSe$_2$ respectively. \textbf{e} Energy-level diagram for the frontier orbitals of gas-phase Me$_6$. \textbf{f-h} Energy-level diagrams for the frontier orbitals of Me$_6$ on each of the three substrates demonstrating single electron charge transfer from molecule to substrate on Au(111) and FeCl$_2$, leading to the gas-phase HOMO splitting into SOMO and SUMO orbitals separated in energy by the Coulomb repulsion, $U$, and two electron transfer from molecule to substrate on NbSe$_2$. Here, $\Phi$ represents the respective substrate's work function.}
    \label{fig1:summary}
\end{figure}

Control of the properties of this new class of magnetic materials via their electron population is also highly desirable. In the context of organic and metal-organic materials on surfaces, one unavoidable factor affecting electron population is charge transfer. Depending on the energy level alignment between electronic states of adsorbates and surfaces, electron transfer may occur upon adsorption. For a single-molecule on a metal surface, this depends upon the molecule's ionization potential (IP), electron affinity (EA), and the surface's work function ($\Phi$) as illustrated schematically in Figure \ref{fig1:summary}e-h. 
In particular, charge transfer between adsorbed  molecules and the substrate, governed by their energy-level alignment and interfacial hybridization,  can modify molecular orbital occupations, stabilize charged  molecular states, and profoundly influence the electronic  properties of the molecular assemblies,\cite{Lu2004,ChargeTransfer2013} or induce open-shell character.\cite{pawlak_toward_2025,wang_aza-triangulene_2022,calupitan_emergence_2023,zonda_resolving_2021,esat_quantum_2024,li_spin_2026} Beyond the single-molecule level, substrate-induced charge redistribution
and electronic screening also modify intermolecular interactions,
influencing both the formation and the collective behavior of molecular
assemblies and enabling properties absent in the isolated building
blocks, such as localized magnetic moments and collective magnetic or
conducting phases.\cite{lowe_local_2024,lobo-checa_ferromagnetism_2024,kinikar_atomic-scale_2025,dominguez-celorrio_systematic_2025} Understanding and controlling the interplay between  molecular structure, intermolecular electronic coupling, and molecule--substrate charge transfer is therefore  essential for establishing surface-supported molecular architectures as a versatile platform for engineering low-dimensional quantum materials with tailored  functionalities.

Here, we systematically investigate charge transfer by studying macrocyclic emeraldine (Me$_6$; see chemical structure in Figure 1a) molecules on three different surfaces - metallic Au(111), a ferromagnetic semiconducting monolayer (ML) of FeCl$_2$/Au(111),\cite{aguirre_ferromagnetic_2024,candia_atomic_2026} and a cleaved bulk crystal of superconducting 2H-NbSe$_2$ (herein referred to simply as NbSe$_2$).
Via scanning tunneling microscopy (STM) and spectroscopy (STS), non-contact atomic force microscopy (ncAFM), and Kelvin probe force microscopy (KPFM) supported by density functional theory (DFT) and model Hamiltonian calculations, we show that Me$_6$ molecules donate one electron to the surface via charge transfer on Au(111) and FeCl$_2$, and two electrons via charge transfer on NbSe\textsubscript{2} due to its larger work function (Figure \ref{fig1:summary}b-d). This leads to  half-filling of the gas-phase HOMO on Au(111) and FeCl$_2$ (Figure \ref{fig1:summary}b,e-f), causing it to split into SOMO and SUMO orbitals separated in energy by the Coulomb repulsion, $U$, and to complete depopulation of the gas-phase HOMO on NbSe$_2$. Our work demonstrates the ability to control the macrocycle's spin and charge by energy-level alignment dictated by substrate work function, and provides experimental and theoretical tools for rationalizing and quantifying charge transfer on different surfaces.

\section{Results and discussion}
\subsection{Sample preparation and structural properties}
To compare the properties of Me$_6$ molecules on both Au(111) and FeCl$_2$, we prepared samples with 0.5 ML coverage of FeCl$_2$/Au(111) by sublimating stoichiometric FeCl$_2$ powder onto a clean Au(111) crystal held at $\sim$150\textsuperscript{o}C.\cite{aguirre_ferromagnetic_2024,candia_atomic_2026} We then sublimated the Me$_6$ molecules at $\sim320$\textsuperscript{o}C onto the half-FeCl$_2$, half-Au(111) samples held at slightly below room temperature (RT). To prepare NbSe$_2$ samples, we cleaved the crystal under vacuum conditions and subsequently annealed it to $\sim$100$^\mathrm{o}$C in UHV. We then deposited the Me$_6$ molecules using the same sublimation conditions as above with the NbSe$_2$ substrate held at RT.

\begin{figure}
    \centering
    \includegraphics[width=0.5\linewidth]{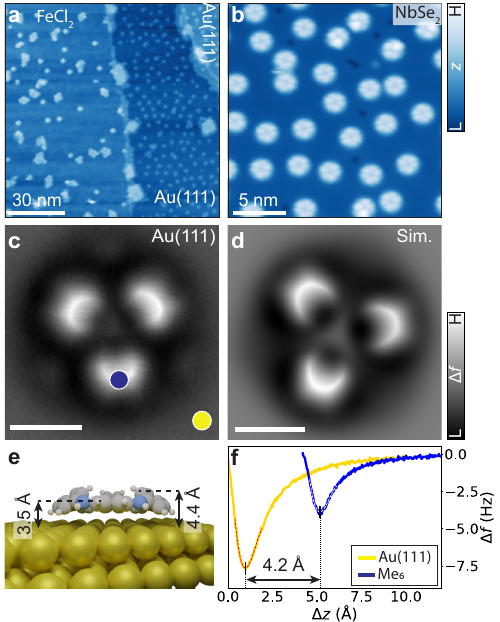}
    \caption{Sample preparations and structure of Me$_6$. \textbf{a,b} Constant-current metal-tip STM overview images showing Me$_6$ molecules adsorbed on terraces of both FeCl$_2$ and Au(111) ($V_\mathrm{b} = 1.6$ V, $I_\mathrm{t} = 3$ pA), and on NbSe$_2$ ($V_\mathrm{b} = 1 $ V, $I_\mathrm{t} = 20$ pA) respectively. 
    \textbf{c} Constant-height CO-tip ncAFM image of an isolated Me$_6$ molecule on Au(111). 
    \textbf{d} Simulation of ncAFM image of Me$_6$ based on DFT optimized structure of Me$_6$/Au(111) using the ProbeParticle method. Panels c,d scalebars: 5 {\AA}.
    \textbf{e} Side-view of DFT optimized structure of Me$_6$/Au(111).
    \textbf{f} Experimental measurement of Me$_6$ adsorption height on Au(111) based on $\Delta f(\Delta z)$ spectra acquired on Me$_6$ (blue) and bare Au(111) (yellow), respectively, at the sites shown in panel c. Dashed lines indicate cubic fits used to accurately determine $\Delta z$ corresponding to $\Delta f_\mathrm{min}$.}
    \label{fig2:struc}
\end{figure}

After preparation, the samples were characterized using STM. Typical STM overview images of the two samples are shown in Figure \ref{fig2:struc}a,b. In panel a, the left half of the image shows a ML FeCl$_2$ terrace and the right half of the image shows two bare Au(111) terraces. The molecules preferentially adsorbed on the Au(111) regions, but at the slightly cool substrate temperature used for preparation some molecules can also be found on the FeCl$_2$ terraces. For the most part, the molecules remain isolated from one another after adsorption (with some molecular clusters also visible on the FeCl$_2$ terrace). Similarly in panel b, the molecules appear to adsorb and remain isolated from one another on the NbSe$_2$ substrate. Herein, we focus exclusively on the properties of isolated molecules.  

On the Au(111) and FeCl$_2$ sample, we prepared a CO-terminated tip and performed ncAFM measurements on Me$_6$ molecules on both substrates. The image on Au(111) is shown in Figure \ref{fig2:struc}c. The molecule has a three-fold symmetric appearance with greatest intensity at three half-ring features and weaker intensity at the center of the molecule. Using DFT, we calculated the optimized structure of Me$_6$ molecules on Au(111). A side view of the optimized structure on Au(111) is shown in Figure \ref{fig2:struc}e. The molecules adopt a non-planar adsorption configuration with three of the six benzene rings of the molecule tilted out of plane making the molecule three-fold symmetric. Based on the optimized geometry on Au(111),  we used the ProbeParticle model to predict the appearance of the Me$_6$ molecules in ncAFM images.\cite{krejci_principles_2017} As shown in Figure \ref{fig2:struc}d, the simulated image strongly resembles the experimental image. Further agreement between theory and experiment was obtained by experimentally measuring the molecule's adsorption height via CO-tip $\Delta f(\Delta z)$ spectroscopy - as shown in Figure \ref{fig2:struc}f. By comparing the relative height of the $\Delta f(\Delta z)$ minimum on Au(111) and on the molecule, we estimate that the tilted molecular ring is 4.2 \AA{} above the Au(111) surface, closely matching the calculated value of 4.4 \AA{} (panel e). 

Similar DFT calculations and CO-tip measurements were performed for Me$_6$ molecules on FeCl$_2$. Here, qualitatively identical ncAFM images were obtained (see Figure S3) and a slightly larger adsorption height was measured - in agreement with DFT calculations (Figure S1). On NbSe$_2$, we were unable to prepare a CO-tip but similar appearance (though lower quality) was observed in an ncAFM image acquired with a Se-passivated tip (see Figure S4). DFT also confirms a similar adsorption height for the molecule in this case (Figure S1). A comparison of experimental measurements with DFT calculations therefore suggests similar structural properties across the three surfaces with Me$_6$ molecules physisorbed in each case.

\subsection{Electronic properties of Me$_6$}
The DFT calculations of the molecule on each of the three substrates also provided a first hint about their electronic properties. By calculating the projected molecular density of states (DOS) of the molecules on each of the three substrates, we found that the gas-phase highest-occupied molecular orbital (HOMO) of Me$_6$ appears to be aligned with the surface Fermi level for Au(111) and FeCl$_2$, and slightly to the unoccupied side of the Fermi level for NbSe$_2$ (Figure S5). These calculations therefore suggest the possibility of charge transfer from molecule to substrate.

\begin{figure}
    \centering
    \includegraphics[width=\linewidth]{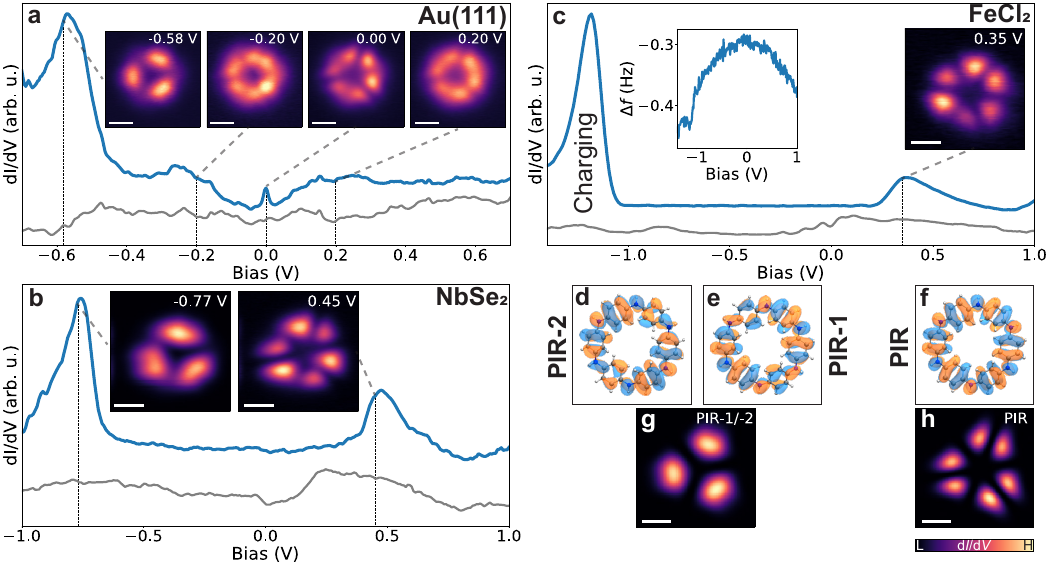}
    \caption{Differential conductance STS measurements of isolated Me$_6$ molecules on Au(111), FeCl$_2$, and NbSe$_2$. \textbf{a-c} STS measurements acquired on isolated Me$_6$ molecules (blue curves) and substrate references (grey curves) on Au(111), NbSe$_2$, and FeCl$_2$, respectively. Setpoints: $V_\mathrm{b} = 700$ mV, $I_\mathrm{t} = 400$ pA (a), $V_\mathrm{b} = 1.4$ V, $I_\mathrm{t} = 100$ pA (b), $V_\mathrm{b} = -1.4$ V, $I_\mathrm{t} = 340$ pA (c) .  Panel c, left inset: KPFM $\Delta f(V_\mathrm{b})$ spectrum simultaneously acquired with the same panel's d$I$/d$V$ spectrum.  All other insets: Constant-height metal-tip d$I$/d$V$ maps acquired on Me$_6$ molecules at the specified bias voltages on each surface. 
    \textbf{d-f} Highest occupied molecular orbitals (HOMOs) of Me$_6$ corresponding to positive ion resonances (PIRs), calculated in the gas-phase using the optimized geometry of Me$_6$ on Au(111) (isosurface: 0.02 $e$\AA{}$^{-3}$). The PIR-1 and PIR-2 are degenerate. 
    \textbf{g-h} Simulated $s$-tip d$I$/d$V$ maps of the first PIRs of gas-phase Me$_6$ with on-surface optimized geometry associated with the orbitals shown in d-f. All scalebars: 5 {\AA}.}
    \label{fig3:STS}
\end{figure}

To test for the possibility of charge transfer, we experimentally probed the electronic properties of the Me$_6$ molecules on each substrate via d$I$/d$V$ STS. A typical d$I$/d$V$ spectrum for an isolated Me$_6$ molecule on Au(111) is shown in Figure \ref{fig3:STS}a. This curve features four distinct peaks - one large peak at $\sim -$0.6 V, two shoulders at $\sim\pm$0.2 V, and a narrow resonance at zero bias (shown in higher resolution in Figure S6). We examined the spatial distribution of these four features by acquiring d$I$/d$V$ maps at each bias of interest, as shown in the insets of Figure \ref{fig3:STS}a. The feature at $\sim$-0.6 V exhibited three homogeneous rectangular lobes across the Me$_6$ molecule whereas the three features closest to the Fermi energy (-0.2, 0, and 0.2 V) all shared a similar appearance featuring six smaller circular lobes. 

For Me$_6$ molecules on NbSe$_2$, d$I$/d$V$ spectra displayed two pronounced peaks, as shown in Figure \ref{fig3:STS}b. The spatial distribution of these two features are shown in the d$I$/d$V$ maps in the insets of Figure \ref{fig3:STS}b. The feature at $\sim-$0.8 V displays three rectangular lobes and the feature at $\sim0.5$ V displays six smaller circular lobes - similar to the two distinct appearances observed on Au(111). 

Finally, for Me$_6$ molecules on FeCl$_2$, d$I$/d$V$ spectra featured two pronounced peaks. The first peak was a tall narrow resonance at $\sim$-1.2 V, and the second a shorter peak at $\sim$0.35 V as shown in Figure \ref{fig3:STS}c. The spatial resolution of the feature at 0.35 V shown in the d$I$/d$V$ map in the right inset of panel c is reminiscent of the near-Fermi character of the features of Me$_6$ molecules on Au(111), with six circular lobes visible. As shown in the left inset of this panel, the tall resonance at $\sim$-1.2 V is accompanied by a discontinuity in the simultaneously acquired KPFM parabola measuring $\Delta f(V_\mathrm{b})$. From this observation,\cite{gross_measuring_2009,stomp_detection_2005} combined with ring-like intensity in d$I$/d$V$ maps revealed in the vicinity of this feature (see Figure S7)\cite{portner_charge_2020,pawlak_toward_2025,yan_synthesis_2021,kumar_electric_2019} and linear shifting of the d$I$/d$V$ peak with varying tip-sample distance(Figure S8)\cite{lowe_local_2024,kumar_electric_2019}, we conclude that this feature is not an intrinsic molecular state but instead a charging feature that arises due to the double-barrier tunnel junction (where the first barrier is the usual tip-sample vacuum barrier, and the second created by the insulating FeCl$_2$ region which separates the Me$_6$ molecule from the underlying metal surface - discussed in greater detail in Section S5).

To interpret the STS results of Figure \ref{fig3:STS}, we calculated the expected appearance of different d$I$/d$V$ features of Me$_6$ molecules. To achieve this, we used DFT to calculate the molecular orbitals of gas-phase Me$_6$ (but with the optimized on-surface geometry), and used these to simulate d$I$/d$V$ maps associated with ion resonances. Note that we recently showed that direct comparison with DFT-calculated molecular orbitals can be potentially unreliable,\cite{kumar_multireference_2025} but in this case these maps are qualitatively identical to calculated Dyson orbitals\cite{ortiz_dyson-orbital_2020} (see comparison in Section S6). Molecular orbitals corresponding to the first positive ion resonances (PIR; occupied states) of gas-phase Me$_6$ are shown in Figure \ref{fig3:STS}d-f, with the PIR-1 and PIR-2 orbitals found to be degenerate. Based on simulated d$I$/d$V$ maps using orbitals, we find that the gas-phase first PIR should exhibit six circular lobes, whereas the PIR-1/-2 should exhibit three larger rectangular lobes, as shown in Figure \ref{fig3:STS}g,h. 

The simulated gas-phase PIR strongly resembles the first negative ion resonance (NIR; unoccupied state) observed at positive bias experimentally on each of the three surfaces. This provides strong evidence of charge transfer from molecule to substrate in each case. On Au(111), the -0.20 V and 0.20 V maps in Figure \ref{fig3:STS}a both exhibit six circular lobes of intensity which resemble the gas-phase PIR in panel h, and the -0.58 V map in panel a resembles the gas-phase PIR-1/-2 in panel g. The observation of features with similar spatial intensity above and below the Fermi level points to a molecular charge of +1 (single electron charge transfer from molecule to surface), leading to a splitting of the gas-phase HOMO into SOMO and SUMO states, separated in energy by Coulomb repulsion, $U$ (as illustrated in Figure \ref{fig1:summary}f). This hypothesis also naturally explains the additional sharp peak observed at zero bias in Figure \ref{fig3:STS}a with a similar spatial distribution observed in the corresponding d$I$/d$V$ map to the -0.20 V and 0.20 V maps. The half-filled HOMO leads to the molecular hosting an unpaired electron and hence a spin 1/2. On the metallic Au(111) surface, this spin can be Kondo screened by the itinerant substrate electrons to produce the well-established signature of a zero-bias peak observed in Figure \ref{fig3:STS}a.\cite{ternes_spectroscopic_2008} Using a Fano function convolved with the Fermi-Dirac distribution to fit this zero bias peak (see higher resolution spectrum in Figure S6),\cite{nagaoka_temperature_2002} we estimate a Kondo temperature of $T_\mathrm{K} = 14.5$ K from its half-width at half-maximum. 

Similarly, there is a strong resemblance between the simulated d$I$/d$V$ maps and the NbSe$_2$ experimental data. The -0.77 V map in Figure \ref{fig3:STS}b resembles the gas-phase PIR-1/-2 in panel g, and the 0.45 V map in panel b resembles the gas-phase PIR in panel h. The observation of the gas-phase PIR feature exclusively at positive bias in this case points to a larger Me$_6$ molecular charge of +2 on NbSe$_2$, with the gas-phase HOMO becoming completely empty as a result of two-electron charge transfer from molecule to substrate (as illustrated in Figure \ref{fig1:summary}h).

For Me$_6$ on FeCl$_2$, the observed state at 0.35 V in Figure \ref{fig3:STS}c has spatial distribution resembling the gas-phase PIR, pointing to charge transfer from molecule to substrate. Because of the charging peak, however, no state could be observed at negative bias which prevents confident attribution of the molecular charge state as either +1 or +2. 

From comparison of simulated and experimental d$I$/d$V$ maps, we attribute the molecular charge state of Me$_6$ on Au(111) as +1, and of Me$_6$ on NbSe$_2$ as +2 but the case of Me$_6$ on FeCl$_2$ remains ambiguous. 

\subsection{Substrate work function and molecular charge}

The origin of the different molecular charge states of Me$_6$ on the different substrates must be related to charge transfer between molecule and respective substrate. One of the most important substrate properties which governs the magnitude of this charge transfer is its work function, as illustrated in Figure \ref{fig1:summary}f-h. The work function of Au(111) and NbSe$_2$ are known to be $\sim$5.25\cite{rusu_surface_2006,hansson_photoemission_1978} and 5.9 eV,\cite{shimada_work_1994} respectively. The larger work function of NbSe$_2$ than Au(111) is consistent with the hypothesis of greater electron transfer from molecule to substrate leading to a molecular charge of +2 on this surface (compare Figure \ref{fig1:summary}f and h). 

\begin{figure}
    \centering
    \includegraphics[width=\linewidth]{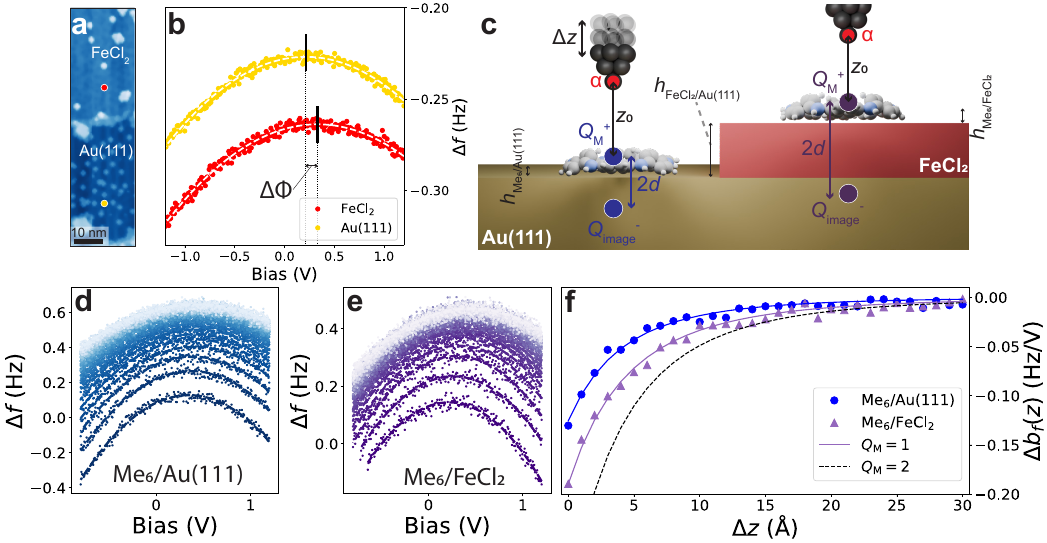}
    \caption{Electrostatic comparison of Au(111) and FeCl$_2$. \textbf{a} Constant-current STM image showing FeCl$_2$ and Au(111) ($V_\mathrm{b} = 1.6$ V, $I_\mathrm{t} = 3$ pA). \textbf{b} CO-tip KPFM measurements acquired on Au(111) and FeCl$_2$ at the sites indicated by colored markers in a. White dashed lines indicate parabolic fits from which the maxima, $V_\mathrm{CPD}$, were extracted, indicated by dotted vertical lines, allowing the difference in work function ($\Delta\Phi$) to be determined. 
    \textbf{c} Illustration of the image charge effect for Me$_6$ on Au(111), and on FeCl$_2$/Au(111) showing adsorption heights, $h$, screening distances, $d$, molecular charges, $Q_\mathrm{M}$, image charges, $Q_\mathrm{image}$, tip sample distance, $z_0 + \Delta z$, and the model parameter associated with charge distribution on the tip, $\alpha$, for the two different scenarios. 
    \textbf{d-e} Tip-height dependent CO-tip KPFM measurements acquired on isolated Me$_6$ molecules on Au(111) and FeCl$_2$. Dashed lines indicate parabolic fits from which the maxima, $V_\mathrm{CPD}$, were extracted. Progressively lighter colors represent progressively larger tip-sample distances. 
    \textbf{f} Plot showing extracted $\Delta b_f$ for Me$_6$/Au(111) and Me$_6$/FeCl$_2$ using Eqs. \eqref{del_f_abc} and \eqref{Eq:b(qm,d)} (more details in Section S7). Solid lines indicate fits to Eq. \eqref{Eq:b(qm,d)} from which $\alpha$ and $z_0$ were determined from the blue fit to the Me$_6$/Au(111), and these parameters used to provide an estimate of molecular charge $Q_\mathrm{M}$ for Me$_6$/FeCl$_2$ from the purple and black fits.}
    \label{fig4:WF}
\end{figure}

To estimate the work function, $\Phi$, of FeCl$_2$ we performed further KPFM measurements. The contact potential difference, $V_\mathrm{CPD}$, obtained from the maximum of $\Delta f(V_\mathrm{b})$ parabolas acquired in KPFM measurements is a measure of the difference between tip and sample work functions ($\Phi_\mathrm{t}$ and $\Phi_\mathrm{s}$ respectively):

\begin{equation}
    V_\mathrm{CPD} = \frac{\Phi_\mathrm{s}-\Phi_\mathrm{t}}{e},
\end{equation}

where $e$ is the charge of an electron. By measuring $V_\mathrm{CPD}$ on Au(111) and on FeCl$_2$ using the same tip, we can therefore estimate $\Phi_\mathrm{FeCl_2}$. KPFM measurements of FeCl$_2$ and Au(111) acquired using the same tip are shown in Figure \ref{fig4:WF}a-b. From these measurements, we find a value of $V_\mathrm{CPD}$ on FeCl$_2$ that is $\sim100$ mV larger than on Au(111). This implies $\Phi_\mathrm{FeCl_2} = \Phi_\mathrm{Au(111)} + 0.1 \approx 5.35$ eV (although we found some evidence for variations in this value across the surface - see discussion in Section S8). Therefore the FeCl$_2$ substrate has a work function in between those of Au(111) and NbSe$_2$. 

To gain greater insight into the charge properties of Me$_6$ molecules on FeCl$_2$, we also performed further KPFM measurements on the molecules on Au(111) and FeCl$_2$ over a range of tip-sample distances, $\Delta z$, using the same tip condition. These measurements revealed a downward trend of $V_\mathrm{CPD}$ values for both Me$_6$/Au(111) and Me$_6$/FeCl$_2$ with decreasing $\Delta z$, with the trend slightly faster for the molecules on FeCl$_2$ - see Figure S13. This downward trend is a strong indication of positive charge on the molecule on both substrates.\cite{mallada_real-space_2021,Gross2014KPFM} Note that for Me$_6$/NbSe$_2$, we were unable to perform KPFM measurements with the same tip conditions. KPFM measurements acquired with a different tip, however, also show this downward trend that supports the hypothesis of positive molecular charge (see Figure S4). It is tempting to attribute the more rapid decrease of $V_\mathrm{CPD}$ for Me$_6$/FeCl$_2$ compared to Me$_6$/Au(111) to a larger positive charge of the molecule on Me$_6$/FeCl$_2$. It is necessary, however, to also consider the screening environments of both charges. Arguably, the charged Me$_6$ molecules are more strongly screened in direct proximity to metallic Au(111) than when they have an insulating FeCl$_2$ layer between them and the Au(111).

To quantitatively compare the Au(111) and FeCl$_2$ KPFM results, we developed a model which accounts for both the molecular charge and difference in screening lengths. This model is discussed in detail in Section S7. In brief, we measured $\Delta f(V,\Delta z)$ curves like those shown in Figure \ref{fig4:WF}d,e, on Me$_6$/Au(111), Au(111), Me$_6$/FeCl$_2$, and FeCl$_2$ while maintaining the same tip conditions. From the parabolic $\Delta f(V)$ curves at each $\Delta z$ value, we performed fits of the form:
\begin{equation}\label{del_f_abc}
    \Delta f(V) = a_f(V-V_\mathrm{CPD}^\infty)^2+b_f(V-V_\mathrm{CPD}^\infty)+c_f,
\end{equation}
where $a_f,b_f,$ and $c_f$ are fitting parameters and $V_\mathrm{CPD}^\infty$ represents the asymptotic value of the $V_\mathrm{CPD}$ as $\Delta z \rightarrow \infty$ (further details in Section S7). The most important of these fitting parameters is the coefficient of the linear term, $b_f$, which can be related to physical quantities such as the molecular charge, $Q_\mathrm{M}$ (in dimensionless units of elementary charge), and the effective image charge distance, $2d$, via:
\begin{equation}\label{Eq:b(qm,d)}
    \Delta b_f(\Delta z) = b_f(\text{Me}_6/\text{subs.}) - b_f(\text{subs.}) = -\alpha Q_\mathrm{M}\left(\frac{1}{(\Delta z-z_0)^3} - \frac{1}{(\Delta z-z_0+2d)^3} \right),
\end{equation}
where $\alpha$ is a parameter related to the condition of the tip (see Section S7) and $z_0$ represents the tip-sample distance at $\Delta z = 0$. Note that $\Delta b_f$, representing the difference between $b_f$ fit to the measurements on Me$_6$ on a particular surface, $b_f(\text{Me}_6/\text{subs.})$, and the reference measurement on that surface, $b_f(\text{subs.})$, was computed to isolate the influence of the Me$_6$ molecules themselves from contributions from the respective surfaces. On both surfaces, $\Delta z = 0$ was defined by retracting the tip 250 pm from the minimum of the measured $\Delta f(\Delta z)$ spectrum on an Me$_6$ molecule on that surface (Figure S13). This model is illustrated schematically in Figure \ref{fig4:WF}c.

Given that we could confidently attribute the molecular charge of Me$_6$/Au(111) via the aforementioned d$I$/d$V$ measurements, our approach was to first fit the data acquired on Au(111) to Eq. \eqref{Eq:b(qm,d)} assigning a molecular charge of $Q_\mathrm{M}=+1$, and choosing $d$ based on the adsorption height of Me$_6$/Au(111) (3.5 \AA{}, Figure \ref{fig2:struc}e) and the estimated leaking of metallic states into vacuum (1.1 \AA{}): $d = 3.5 - 1.1 = 2.4$ \AA{}.\cite{lam_first-principles_1993,neaton_renormalization_2006} By doing so, we could fit the unknown parameter related to the condition of the tip, $\alpha$, and the tip-sample distance at $\Delta z=0$, $z_0$. The resultant fit is the blue curve in Figure \ref{fig4:WF}f. 

Because the same tip was used for the measurements on FeCl$_2$ starting at an equivalent tip-sample distance, we used the same values of $\alpha$ and $z_0$ extracted from the fitted Me$_6$/Au(111) data to then fit the Me$_6$/FeCl$_2$ data using Eq. \eqref{Eq:b(qm,d)} in order to estimate the molecular charge on this surface, $Q_\mathrm{M}$. To do this, we imposed physical upper and lower bounds on the value $d$ (see Section S7), and then performed two fits at fixed values of $Q_\mathrm{M}=+1$ and $Q_\mathrm{M}=+2$. The curve associated with $Q_\mathrm{M} = +1$ is shown in purple in Figure \ref{fig4:WF}f. From this fit, we extract a realistic value of $d = 7.2 \pm 0.4$ \AA{}. By contrast, the fit with $Q_\mathrm{M} = +2$ could not converge and $d$ became saturated at its lower-bound of 3.7 \AA{} (corresponding to the Me$_6$/FeCl$_2$ adsorption height; Figure S1). Alternative methods in which we estimated the value of $d$ and left $Q_\mathrm{M}$ as a fitting parameter also support the claim of a +1 charge of Me$_6$ on this surface (see Figure S15). These KPFM results provide strong evidence that the charge of Me$_6$/FeCl$_2$ is +1, arising from a single electron charge transfer from molecule to substrate. 

\subsection{Modeling charge transfer}

The DFT calculations in Figure S5 provided first theoretical hints about our interpretation of charge transfer. However, this approach has some limitations. Firstly, conventional density functional theory (DFT) calculations  employing local or semilocal exchange--correlation functionals systematically underestimate the fundamental electronic gap  of molecules and solids.\cite{Perdew1983,Sham1983,Hybertsen1986,Onida2002} Secondly, though the DOS peak corresponding to the HOMO appears to be almost entirely shifted to the unoccupied side of the Fermi energy on NbSe$_2$, it remains somewhat pinned to the Fermi energy, making it very difficult to quantify the charge transfer. Additionally, DFT poorly captures the bandgap of FeCl$_2$ (see Section S3) relative to experimental observations,\cite{aguirre_ferromagnetic_2024,candia_atomic_2026} possibly resulting in physically unrealistic charge transfer directly between Me$_6$ and FeCl$_2$ rather than with the underlying Au(111). Finally, though not unprecedented,\cite{ChargeTransfer2013,dominguez-celorrio_systematic_2025,biswas_-surface_2021,jiang_many-body_2023} the two electron charge transfer from molecule to substrate on NbSe$_2$ is somewhat surprising given the modest work function difference of $\sim$0.65 eV between Au(111) and NbSe$_2$ - a finding that merits further examination. 

To rationalize the influence of the surface on adsorbed Me$_6$ molecules and provide greater insight into the nature of the charge transfer, we employ a model Hamiltonian that incorporates both the static polarization of the molecular electronic structure induced by a conductive substrate, and the dynamic exchange of electrons between molecule and surface: 

\begin{equation}
    \hat{H} = \hat{H}_\text{mol} + \Delta\hat{H}_{\mathrm{im}} + \hat{H}_\text{surf} + \hat{H}_\text{hyb},
\end{equation}

where $\hat{H}_\mathrm{mol}$ is the many-electron Hamiltonian of the isolated, gas phase molecule obtained from quantum chemistry calculations, $\Delta \hat{H}_\mathrm{im}$ describes the renormalization of the molecular electronic structure due to image-charge screening by a conductive surface, $\hat{H}_\mathrm{surf}$ represents the electronic states of the surface, and $\hat{H}_\mathrm{hyb}$ accounts for electronic coupling between molecule and surface. See Methods for further details of each of these terms. 

\begin{figure}
    \centering
    \includegraphics[width=1.0\linewidth]{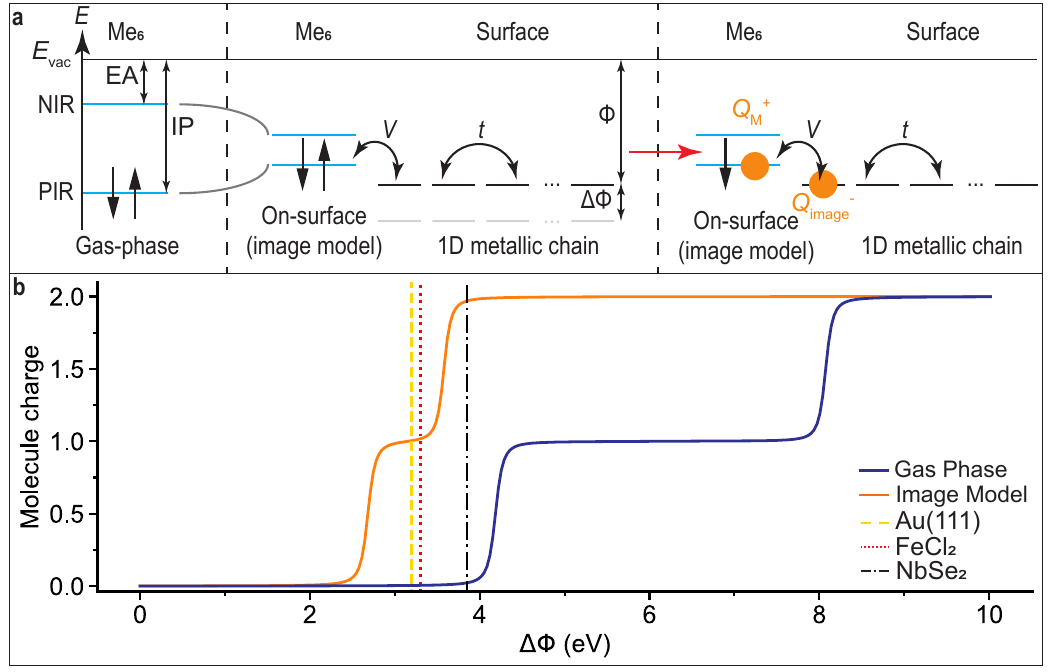}
    \caption{Modelling the influence of a substrate on Me$_6$. 
    \textbf{a} Illustration of the model Hamiltonian. The image model causes renormalization of the frontier orbitals due to the image charge effect. The coupling between the frontier orbitals of Me$_6$ to a conductive surface is approximated by a 1D metallic chain. $V$ represents molecule-surface hybridization, $t$ electron hopping within the 1D metallic chain, and $\Phi$ surface work function. At values of $\Phi$ which promote molecule to surface charge transfer, the charge of the molecule is screened by its image charge in the surface.
    \textbf{b} Molecular charge of Me$_6$ as a function of $\Delta\Phi$ in both the gas-phase and with image-charge renormalization included. $\Delta\Phi = 0$ is defined as a situation in which the Fermi energy is exactly halfway between the PIR and NIR. Vertical lines indicate estimated positions of the work function of each of the three surfaces.}
    \label{fig5:models}
\end{figure}

Within this framework, the first term describes the molecular states of Me$_6$ and the second is responsible for the renormalization of these states in the presence of a conductive surface. As illustrated in Figure \ref{fig5:models}a, fluctuations in charge density between an Me$_6$ molecule in proximity with a conductive surface induce mirror charges in the surface. The electrostatic interaction between the molecular charge and its induced image charge causes renormalization of its electronic spectrum: effectively lowering the energy required to add or remove an electron from the molecule and screening electron-electron interactions.\cite{neaton_renormalization_2006,thygesen_renormalization_2009} Within this model, we set half the effective distance between a molecular charge and its image to be 2.4 \AA{}, as was the case for the KPFM model of Me$_6$/Au(111). 

The second two terms in the model Hamiltonian then describe a conductive surface and its coupling with a molecule. As shown schematically in Figure \ref{fig5:models}a, we simplify the conductive surface as a 1D metallic tight-binding chain (see Methods).\cite{zuzak_-surface_2024} Using this full model Hamiltonian, we are able to quantitatively describe changes in Me$_6$'s molecular charge as a function of change in substrate work function. As can be seen in the blue curve Figure \ref{fig5:models}b, a charge state of +1 is extremely stable for gas-phase Me$_6$ molecules, requiring a change of work function of greater than 3 eV before a second electron can be removed. When image charge corrections are included, however, the charge +1 plateau becomes much shorter with only $\sim$0.5 eV change in work function required to remove a second electron. This model therefore provides strong quantitative agreement with the experimental findings on Au(111) and NbSe$_2$, where a change in work function of $\sim$0.65 eV was enough to transition from a charge of +1 to a charge of +2. This theory shows the crucial role of image charge-induced electronic renormalization in governing charge transfer processes. 

The small change in work function of only $\sim$0.1 eV between Au(111) and FeCl$_2$ also supports the attribution of a charge state of +1 on both surfaces within our model Hamiltonian. Although our model does not capture the reduction in the image charge effect likely to be caused by the insulating FeCl$_2$ monolayer, as alluded to in the previous section via a larger $d$, we suggest that such a reduction in the image charge effect is likely to push the model towards the gas phase curve - making a two electron charge transfer scenario even less likely for FeCl$_2$. 

Throughout this work we are talking about integer electron charge transfer and orbital occupation. This is appropriate in the limit of weak physisorption, which the optimized structures and measured adsorption heights (Sections S1-S2) suggest is the case for Me$_6$ on all three surfaces. Within the model Hamiltonian, this is represented as a relatively small hopping term in $\hat{H}_\mathrm{hyb}$ that results in steep transitions between charge states as shown in Figure \ref{fig5:models}b. A larger molecule-substrate interaction would reduce the gradient of the transition between charge states and potentially allow for fractional occupation in the case of a strongly hybridized molecule-substrate systems. 

The KPFM measurements and the model Hamiltonian approach both suggest a single electron charge transfer of Me$_6$ to FeCl$_2$/Au(111). As in the case of Me$_6$/Au(111), this implies the presence of an unpaired electron in the molecule on FeCl$_2$, and hence a local magnetic moment. On Au(111), we were able to directly observe a manifestation of this local magnetic moment via the presence of a Kondo peak. On FeCl$_2$, however, we were unable to observe evidence of magnetism. While this is beyond the scope of this work, the interaction of this local moment with the ferromagnetic FeCl$_2$ surface may be an interesting avenue for future investigation.

\section{Conclusion}

In this work, we have systematically compared the properties of a single molecule on a metal surface, an insulating ferromagnetic surface, and a superconducting surface, providing insight into the role of charge transfer and screening in governing molecular properties. This work provides experimental and theoretical tools for quantifying and rationalizing charge transfer, and comparing adsorbate properties across different surfaces. Namely, we established methodology for quantifying molecular charge states and demonstrate the crucial role of image-charge screening in determining electron transfer at molecule–surface interfaces employing Kelvin Probe force spectroscopy. We show that a relatively small change in substrate work function ($\sim$0.65 eV) can induce a transition between distinct integer molecular charge states, providing a route toward engineering molecular electronic and magnetic properties through substrate selection. This work will be valuable as molecular systems are increasingly prepared on more diverse surfaces of both fundamental and technological interest. 

\section{Methods}
\subsection{Samples preparation}
The Au(111) crystal was prepared by several cycles of sputtering with Ar$^+$ ions and subsequent annealing at $\sim$500 $^\mathrm{o}$C in ultra-high vacuum (UHV). The 2D FeCl$_2$ was prepared using stoichiometric powder commercially obtained from Sigma-Aldrich with a purity of 98\%. The powder was sublimated in UHV using a Knudsen cell at $\sim$430 $^\mathrm{o}$C for 90 seconds onto the Au(111) sample held at $\sim$150 $^\mathrm{o}$C, resulting in a coverage of approximately 0.5 ML. The Me$_6$ molecules were then sublimated in UHV using a Knudsen cell at $\sim$320 $^\mathrm{o}$C for 60 seconds with the sample cooled slightly below room temperature (RT), resulting in a coverage of $<$0.1 ML. 

The 2H-NbSe$_2$ crystal was cleaved in vacuum and then annealed in UHV to $\sim$100 $^\mathrm{o}$C. The Me$_6$ molecules were then sublimated using the same conditions as above with the NbSe$_2$ crystal kept at RT during deposition. 

\subsection{Scanning probe microscopy measurements}
The majority of scanning probe microscopy (SPM) measurements were performed in a SPECS-JT microscope at a measuring temperature of 4.4 K using a Kolibri sensor (f$_0 \sim$1 MHz, Q $\sim$160k, $k\sim$1080 kN/m, 50 pm amplitude modulation). NcAFM measurements were performed by first functionalizing the tip with a carbon-monoxide (CO) molecule unless otherwise specified. KPFM measurements in Figure \ref{fig4:WF}b were performed with the same Kolibri sensor functionalized with CO using an oscillation amplitude of 200 pm. These measurements were taken at a height defined by 1.1 nm above the minimum of a $\Delta f(\Delta z)$ measurement performed on FeCl$_2$. A lock-in method was used for both STS and d$I$/d$V$ maps using V$_\mathrm{mod} = $ 5 mV, f$_\mathrm{mod} = $ 723 Hz.

The constant-height ncAFM image in Figure \ref{fig2:struc}c, and KPFM measurements in Figure \ref{fig4:WF}d-f and Section S7 were performed in a Createc LT-STM at a measuring temperature of 4.4 K using a qPlus sensor (f$_0 \sim$ 30 kHz, Q $\sim$80k, $k\sim$1.8 kN/m, 50 pm amplitude modulation). The ncAFM and KPFM measurements were performed with a CO-functionalized tip. For the KPFM measurements in Figure \ref{fig4:WF}d-f (and Section S7), an oscillation amplitude of 200 pm was used and the tip-sample distance corresponding to $\Delta z = 0$ was defined by retracting the tip 250 pm above the minimum of a $\Delta f(\Delta z)$ measurement performed on an Me$_6$ molecule on the respective substrate (see Figure S13). 

\subsection{DFT and Quantum Chemistry Calculations}
The geometries of all molecules were optimized in their respective ground states using density functional theory (DFT) as implemented in the FHI-AIMS software package.\cite{blum_ab_2009} The hybrid PBE0 functional\cite{adamo_toward_1999} was employed for these calculations, with the Tkatchenko-Scheffler \cite{tkatchenko_accurate_2009} method incorporated to account for van der Waals interactions. To construct the molecular Hamiltonians employed in the models,  we generate the one- and two-body integrals using the ORCA quantum chemistry software,\cite{neese_orca_2012} based on the closed-shell DFT orbitals obtained with the PBE0 exchange-correlation functional.\cite{adamo_toward_1999} The model of the molecular Hamiltonian coupled to a discrete chain is solved at DMRG level of theory with the iTensor package in Julia, which provides a tensor network-based implementation of DMRG suitable for lattice models.

\subsection{Model Hamiltonian}
\subsubsection*{Metal–Molecule Interface}

To describe charge transfer at a metal–molecule interface, we employ an effective Hamiltonian that incorporates both the static polarization of the molecular electronic structure induced by the metal substrate and the dynamic exchange of electrons between the molecule and the metallic electrode. The total Hamiltonian is written as

\begin{equation}
\hat{H} = \hat{H}_{\mathrm{mol}} + \Delta\hat{H}_{\mathrm{im}} + \hat{H}_{\mathrm{metal}} + \hat{H}_{\mathrm{hyb}},
\label{eq:total_hamiltonian_interface}
\end{equation}

where $\hat{H}_{\mathrm{mol}}$ is the many-electron Hamiltonian of the isolated molecule obtained from quantum chemistry calculations, $\Delta\hat{H}_{\mathrm{im}}$ describes the renormalization of the molecular electronic structure due to image-charge screening by the metal, $\hat{H}_{\mathrm{metal}}$ represents the electronic states of the metallic electrode, and $\hat{H}_{\mathrm{hyb}}$ accounts for the electronic coupling between the molecule and the metal.

Within this framework, the first two terms determine the effective electronic structure of the adsorbed molecule, while the latter two govern charge exchange across the interface.

\subsubsection*{Image-Charge Renormalization}

When a molecule is brought into proximity with a metallic surface, fluctuations in its charge density induce mirror charges in the metal. The electrostatic interaction between the molecular charge and its induced image charge lowers the energy required to add or remove an electron from the molecule, leading to a renormalization of its electronic spectrum. As a consequence, occupied molecular orbitals are shifted upward in energy, unoccupied orbitals are shifted downward, and the HOMO–LUMO gap is reduced relative to the gas-phase molecule.

This effect is incorporated through the image-charge correction

\begin{equation}
\Delta\hat{H}_{\mathrm{im}} = -\frac{1}{2}V_{\mathrm{im}} (\hat{N}-N_0)^2,
\label{eq:image_correction_interface}
\end{equation}

where $\hat{N}$ is the molecular particle-number operator, $N_0$ is the electron number of the neutral molecule, and $V_{\mathrm{im}}$ denotes the image-charge interaction strength, which depends on the molecule–surface separation and the dielectric response of the substrate. After normal ordering, this correction modifies both the one-electron and two-electron components of the molecular Hamiltonian, thereby accounting for the screening of electron–electron interactions by the metallic environment. We set $V_{\mathrm{im}} = 1/(4|z-z_0|)$ using the adsorption height of the molecule $z$ and the image plane position $z_0$ to match the experimental findings on Au(111) such that $|z-z_0| = 2.4$ \AA{}.\cite{neaton_renormalization_2006} 

\subsubsection*{Metal Hamiltonian and Hybridization}

Following the image-charge renormalization, the molecule is coupled to the metallic electrode to describe charge transfer across the interface. The metal is represented by a one-dimensional tight-binding chain,

\begin{equation}
\hat{H}_{\mathrm{metal}} = \sum_i \varepsilon_i \hat{c}_i^\dagger \hat{c}_i - t \sum_{\langle i,j\rangle} \left( \hat{c}_i^\dagger\hat{c}_j + \hat{c}_j^\dagger\hat{c}_i \right),
\label{eq:metal_hamiltonian_interface}
\end{equation}

where $\hat{c}_i^\dagger$ ($\hat{c}_i$) creates (annihilates) an electron on site $i$, $\varepsilon_i$ is the on-site energy, and $t$ is the nearest-neighbor hopping parameter that determines the bandwidth of the metallic states. We used $t = 1$ eV and varied $\epsilon_i$ to create effective changes in work function, $\Delta \Phi$, as illustrated in Figure \ref{fig5:models}.

The interaction between the molecule and the metal is described by the hybridization Hamiltonian,

\begin{equation}
\hat{H}_{\mathrm{hyb}} = \sum_{m,i} V_{mi} \left( \hat{a}_m^\dagger\hat{c}_i + \hat{c}_i^\dagger\hat{a}_m \right),
\label{eq:hybridization_interface}
\end{equation}

where $V_{mi}$ denotes the coupling between molecular orbital $m$ and metal site $i$. This term enables electron tunneling between the molecule and the electrode and is responsible for charge redistribution, level broadening, and partial occupation of molecular orbitals. We used $V_{mi} = 0.1$ eV. 

The complete Hamiltonian therefore separates the two dominant physical processes governing metal–molecule interfaces. The image-charge term provides a static renormalization of the molecular electronic structure arising from polarization of the metallic substrate, whereas the hybridization term describes the dynamic exchange of electrons between the molecule and the metal. This decomposition allows the influence of electrostatic screening and electronic coupling on charge transfer to be analyzed independently while remaining within a unified many-body framework.

Vertical lines in Figure \ref{fig5:models} indicate estimated positions of the work function of each of the three surfaces. Given the confident attribution of a charge state of +1 for Me$_6$/Au(111) from d$I$/d$V$ measurements, we chose the Au(111) vertical line to be aligned with the center of the +1 charge plateau. Then, using the known and measured work function differences between Au(111), FeCl$_2$, and NbSe$_2$, we added the lines for these two other surfaces.

\begin{suppinfo}
Measured and calculated adsorption heights, ncAFM measurements on FeCl$_2$ and NbSe$_2$ surfaces, DFT calculations, supporting STS measurements of Kondo feature on Au(111) and charging signatures on FeCl$_2$, simulated d$I$/d$V$ maps and Dyson orbitals, detailed KPFM model with supporting measurements, evidence for local variations in FeCl$_2$ work function. 
\end{suppinfo}

\section{Acknowledgements}
The authors acknowledge support from GA\v{C}R 25-17866X, GA\v{C}R 23-06781M, GA\v{C}R 23-05486S, and the CzechNanoLab Research Infrastructure, supported by MEYS CR (LM2023051). We also acknowledge financial support from the TERAFIT project (CZ.02.01.01/00/22\_008/0004594). B.L. acknowledges support from GA\v{C}R 25-16632I and from the European Union’s Horizon Europe research and innovation programme under the Marie Skłodowska-Curie grant agreement No. 101203634. Z.S. was supported by ERC-CZ program (project LL2101) from Ministry of Education Youth and Sports (MEYS) and by the project Advanced Functional Nanorobots (reg. No. CZ.02.1.01/0.0/0.0/15\_003/0000444 financed by the EFRR). The authors would also like to thank Dr. Benjamin Mallada and Dr. Alejandro Jim\'enez Mart\'in for experimental support and advice. 


\bibliography{Refs}

\clearpage

\end{document}